\documentclass[%
 reprint,
 amsmath,amssymb,
 twocolumn,
 aps,
 prl,
]{revtex4}

\usepackage{graphicx}
\usepackage{dcolumn}
\usepackage{bm}
\usepackage[utf8]{inputenc}
\usepackage[T1]{fontenc}
\usepackage{mathptmx}
\newcommand{\D}{\mathrm{\Delta}}

\usepackage{color}
\usepackage{soul}

\begin{document}
\preprint{APS/123-QED}

\title{Spin$-1$ Weyl Point and Surface Arc State in a Chiral Phononic Crystal}

\author{Xiaotian Shi}
\affiliation{Aeronautics and Astronautics, University of Washington, Seattle, WA 98195, USA}%
\author{Jinkyu Yang}
\thanks{Corresponding author\\ jkyang@aa.washington.edu}
\affiliation{Aeronautics and Astronautics, University of Washington, Seattle, WA 98195, USA}%


\begin{abstract}

Spin$-1$ Weyl point is formed by three bands touching at a single point in the three dimensional (3D) momentum space, with two of which show cone-like dispersion while the third band is flat. Such triply degenerate point carries higher topological charge $\pm2$ and can be described by a three band Hamiltonian. We first propose a tight-binding model of a 3D Lieb lattice with chiral interlayer coupling to form the Spin-1 Weyl point. Then we design a chiral phononic crystal that carries these spin$-1$ Weyl points and special straight-type acoustic Fermi arcs. We also computationally demonstrate the robust propagation of the topologically protected surface states that can travel around a corner or defect without reflection. Our results pave a new way to manipulate acoustic waves in 3D structures and provide a platform for exploring energy transport properties in 3D spin$-1$ Weyl systems.
\end{abstract}

\maketitle

In the past decade, Weyl semimetals \cite{wan2011,fang2012a,soluyanov2015a} have become a research focus in the field of three dimensional (3D) topological states, which are characterized by the touching of two bands with linear dispersion in all directions of the 3D momentum space, namely the Weyl point. Weyl points behave as monopoles of Berry flux in the reciprocal space, which carry a non-zero topological charge (or Chern number) \cite{fang2003}. Such topological invariants result in the robustness of the Weyl points that are stable to the small perturbations and cannot be easily gapped. Previous research in Weyl materials has demonstrated a variety of exotic phenomena, such as robust surface states \cite{wan2011} and a chiral anomaly \cite{nielsen1983}. In parallel, Weyl points have also been realized in other classic systems of electromagnetic \cite{lu2013,lu2015,chen2016,chang2017a,wang2017}, acoustic~\cite{xiao2015,yang2016,li2018,he2018,ge2018,xie2019}, and stress waves \cite{wang2018,shi2019}, leading to the novel applications, such as negative refraction \cite{he2018} and collimation effect \cite{ge2018}. In addition to the single Weyl point with a topological charge ($\pm1$), double Weyl points carrying higher topological charges ($\pm2$) have also been discovered \cite{fang2012a,chen2016,chang2017a,he2018,wang2018}, which are formed by the degeneracy of two bands with quadratic dispersion in a certain momentum plane. 

Recently, a new type of triply-degenerate point of topological charge ($\pm2$), referred to the spin-1 Weyl point \cite{bradlyn2016}, has started to attract much attention. It is formed by the linear degeneracy of three bands having cone-like dispersion with a flat band located at the touching point. This can be described by a simple three band ${\mathbf{k}} {\cdot} {\mathbf{S}}$ Hamiltonian with the spin-$1$ vector $\mathbf{S}$ \cite{bradlyn2016}. Spin-1 Weyl points have been theoretically predicted in the condensed mater systems \cite{bradlyn2016,tang2017,chang2017,zhang2018}, cold atoms \cite{zhu2017,fulga2018,hu2018} and then verified in real materials \cite{miao2018,takane2019,rao2019,sanchez2019,schroter2019}. Y. Yang \textit{et al}. designed and fabricated the first 3D phononic crystal with space group $P2_13$ (No. $198$) that carries the acoustic spin-$1$ Weyl point~\cite{yang2019}. They experimentally demonstrated the double Fermi arcs and topologically protected negative refraction of the surface states. However, the unit cell employed is fairly complicated with a non-symmorphic structure. Also, the system hosts both charge$-2$ threefold and fourfold degenerate points in the Brillouin zone, where the wave propagation is affected by both kinds of degenerate points. The question naturally arises as to whether one could design a simple structure to study the acoustic wave transport properties in the 3D spin$-1$ Weyl systems.

In this manuscript, we realize the spin$-1$ Weyl points in a 3D acoustic system. We start with a tight binding model for a Lieb lattice with chiral interlayer interaction. Previous studies have shown that a 2D Lieb lattice~\cite{shen2010}, along with other types of 2D lattices such as ${\mathcal{T}}_{3}$ lattice~\cite{bercioux2009} and Kagome lattice~\cite{green2010}, can form spin$-1$ Dirac points. We construct a 3D Weyl structure by stacking up the 2D subsystems with the help of the synthetic gauge flux, which is introduced by the appropriate coupling in the third dimension~\cite{xiao2015}. We verify that this 3D chiral Lieb lattice can support one pair of spin$-1$ Weyl points of topological charge $\pm2$ in the first Brillouin zone. The associated Fermi arcs and topological surface arc states are also demonstrated in the proposed architecture. The nearly straight Fermi arcs indicate that we can realize a collimated and robust propagation of surface waves~\cite{ge2018}.

\begin{figure}[ht]
\includegraphics[width=3.4in]{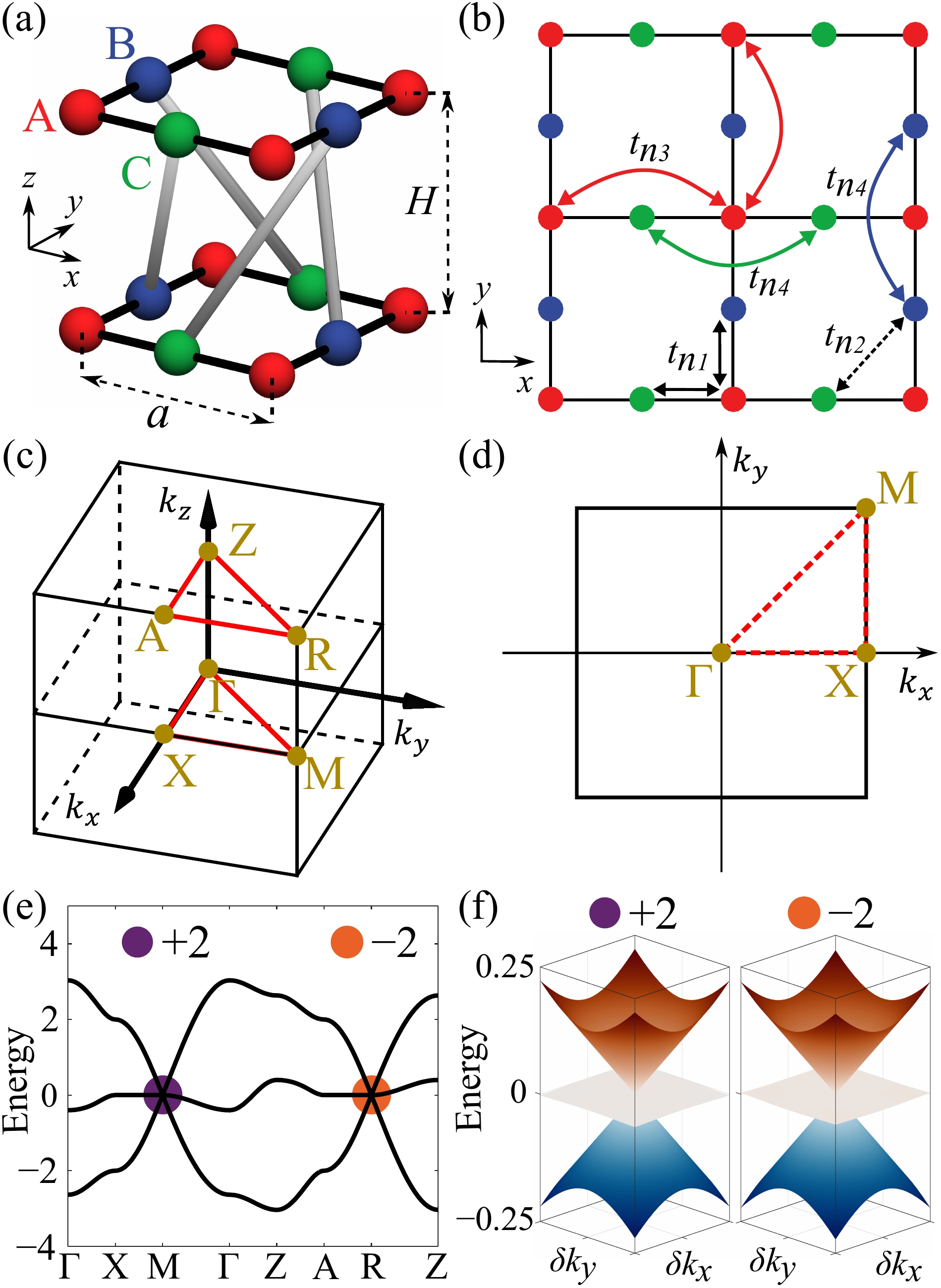}
\caption{(a) Schematic of the 3D Lieb lattice with chiral interlayer couplings. (b) Illustration of in plane short- and long-range intralayer hoppings. (c) First Brillouin zone of the system. (d) 2D reduced reciprocal $k_x-k_y$ plane with fixed $k_z=0$. (e) Band structure of the $3-$band Hamiltonian with purple and orange points indicating the spin-$1$ Weyl points. (f) 2D band structure near the spin-$1$ Weyl points at the $M$ and $R$ points.}
\label{fig:FiG1}
\end{figure}

\begin{figure}[t]
\includegraphics[width=3.4in]{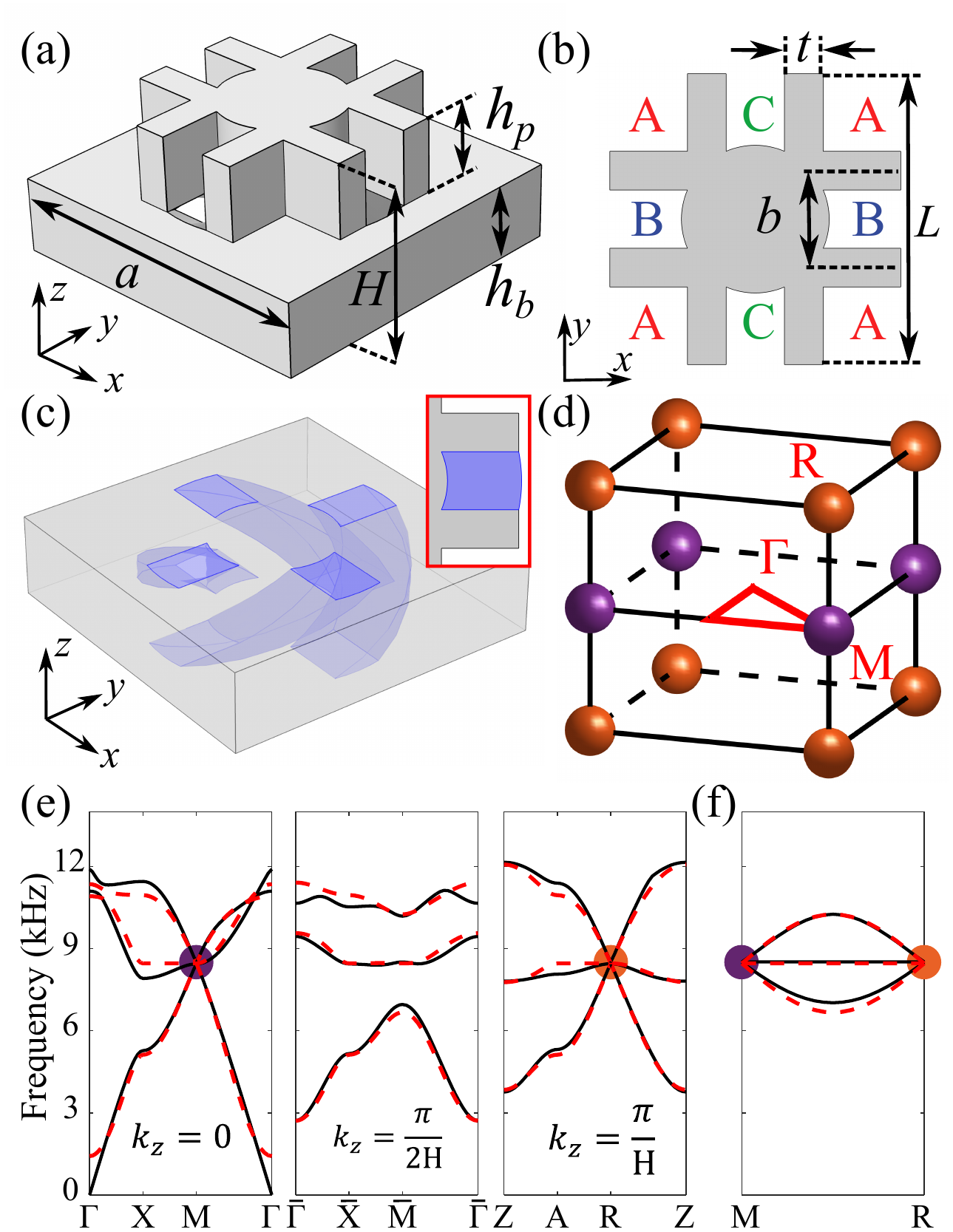}
\caption{(a) Schematic of the unit cell of the chiral phononic crystal. (b) Top view of the upper pillar of the unit cell. (c) Lower perforated plate base of the unit cell. (d) Distribution of the spin-$1$ Weyl points within the first Brillouin zone. (e) Bulk band structure of the phononic crystal in the reduced 2D reciprocal $k_x-k_y$ plane for fixed $k_z=0$, $k_z=\pi/2H$, and $k_z=\pi/H$. (f) Bulk band structure along the $MR$ direction. {The results calculated by  full wave simulation in COMSOL and long-range tight-binding model are shown in (e) and (f) as the solid black lines and dashed red lines, respectively.}}

\label{fig:FiG2}
\end{figure}

We begin with a tight-binding model of the 3D Lieb lattice with chiral interlayer coupling, as shown in Fig. \ref{fig:FiG1}(a). The unit cell has an in-plane lattice constant $a$ and out-of-plane lattice constant $H$, containing three sites A (red sphere), B (blue sphere), and C (green sphere). Therefore, we can write the Hamiltonian of the tight-binding model in the momentum space as:
\begin{equation*}
H(\mathbf{k}) = 
	\begin{bmatrix}
      	\varepsilon_1 + \gamma                & 2t_{n1} \text{cos}(\frac{k_y a}{2})       & 2t_{n1} \text{cos}(\frac{k_x a}{2}) \\
      	2t_{n1} \text{cos}(\frac{k_y a}{2})   & \varepsilon_2 + 2t_{n4} \text{cos}(k_ya)  & \alpha+i\beta \\
       	2t_{n1} \text{cos}(\frac{k_x a}{2})   & \alpha-i\beta                             & \varepsilon_2 + 2t_{n4} \text{cos}(k_xa)
     \end{bmatrix}
\end{equation*}
\noindent {Here, $\varepsilon_1$ stands for the onsite potential on A sites, while $\varepsilon_2$ denotes the onsite potential on B and C sites, $\alpha=[t_{n2}+4t_c\text{cos}(k_z H)] \text{cos}(\frac{k_x a}{2})\text{cos}(\frac{k_ya}{2})$, $\beta=4t_c \text{sin}(k_z H) \text{sin}(\frac{k_x a}{2})\text{sin}(\frac{k_ya}{2})$ and $\gamma = 2t_{n3} [\text{cos}(k_xa) + \text{cos}(k_ya)]$. $t_c$ represents the interlayer hopping while $t_{ni}$ highlight the intralayer hopping. 

Figure {\ref{fig:FiG2}(b)} provides more details of the definition of different intralayer hopping parameters. Specifically, $t_{n1}$ and $t_{n2}$ are the short-range hopping, which refers to the nearest-neighbor (between $A$ and $B/C$ sites) and next-nearest-neighbor (between $B$ and $C$ sites) hopping respectively. In this study, we introduce not only short-, but also long-range hopping to improve the accuracy of the tight-binding model. To this end, we introduce $t_{n3}$ ($t_{n4}$), which stands for the long-range hopping between $A$ sites ($B/C$ sites).  The first Brillouin zone and the reduced 2D reciprocal $k_x-k_y$ plane are given in Figs. {\ref{fig:FiG1}}(c) and (d).} 

{Without loss of generality, we first consider a simplified tight-binding model with only intralayer nearest neighbour hopping $t_{n1}$ and the chiral interlayer hopping $t_c$.} In Fig. \ref{fig:FiG1}(e), we show a typical band structure of the tight-binding model along the high-symmetry lines in the first Brillouin zone with hopping parameters $\varepsilon_1=\varepsilon_2=0$, $t_{n1}=1$, and $t_c=0.1$. Evidently, the three bands degenerate at the $M$ point as indicated by the purple point in Fig. \ref{fig:FiG1}(e). {To investigate the topological property of this triply degenerate point, we expand the Hamiltonian around $M(\frac{\pi}{a},\frac{\pi}{a},0)$ point, which gives:}

\begin{eqnarray*}
\begin{aligned}
H(\D \textbf{k}) = \varepsilon S_0-t_{n1}\D k_x S_1-t_{n1}\D k_y S_2 - 4t_c\D k_z S_3.
\end{aligned}
\end{eqnarray*}

\noindent {Here, $\D \textbf{k}=(\D k_x, \D k_y,\D k_z)$ is a small $k$-vector deviating from the $M$ point, $S_0$ is the $3 \times 3$ unit matrix, and $S_1$, $S_2$, $S_3$ are three of the Gell-Mann matrices given as~}{\cite{beugeling2012}}:

\begin{equation*}
S_1 = 
	\begin{pmatrix}
	    0  & 0  & 1 \\
      	0  & 0  & 0 \\
       	1  & 0  & 0 \\
     \end{pmatrix}
,  
S_2 = 
	\begin{pmatrix}
      	0  & 1  & 0 \\
      	1  & 0  & 0 \\
       	0  & 0  & 0 \\
     \end{pmatrix}
, 
S_3 = 
	\begin{pmatrix}
      	0  & 0  & 0 \\
      	0  & 0  & -i \\
       	0  & i  & 0 \\
     \end{pmatrix}
.
\end{equation*} 
 
\noindent {Such an linearized Hamiltonian describes a spin-1 Weyl points of topological charge +2, which is a natural generalization of the regular weyl point} \cite{beugeling2012,bradlyn2016}. 

Near the degenerate point in Fig. \ref{fig:FiG1}(e), the first and third bands have linear dispersion, while the second band remains nearly flat, which is typical behavior of the spin-$1$ Weyl point. Similarly, there exists another spin-$1$ Weyl point with topological charge $-2$ at the corner of the first Brillouin zone ($R$ point). Figure \ref{fig:FiG1}(f) shows the surface map of the 2D band structure near the two spin-$1$ Weyl points, which gives a better visualization of the Weyl cones that interact with a nearly flat band in gray color. 

\begin{figure}[ht]
\includegraphics[width=3.5in]{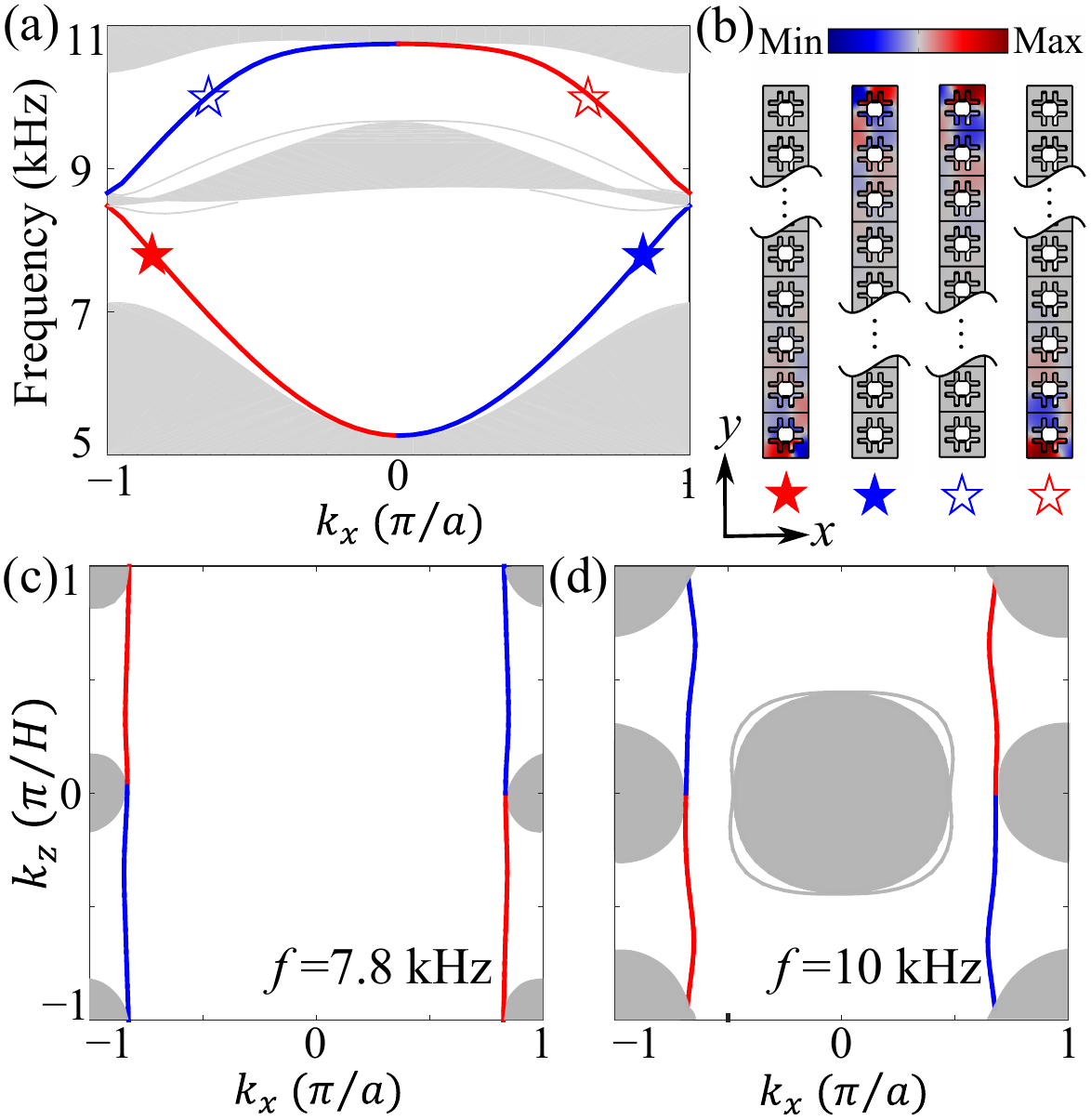}
\caption{(a) Projected band structure of the supercell along the $x$ direction for fixed $k_z=\pi/2H$. (b) Calculated eigenmodes of surface arc states at $7.8$ kHz or $10$ kHz corresponding to the stars in (a). Color intensity represents the magnitude of acoustic pressure field. Equifrequency contours at (c) $f = 7.8$ kHz and (d) $f = 10$ kHz, respectively. The red (blue) lines represent the acoustic Fermi arcs on the negative (positive) $xz$ plane.}
\label{fig:FiG3}
\end{figure}

Although the tight-binding model is just a simple toy model, it provides a keen insight into the physics and a guideline for designing the realistic phononic systems. Inspired by the recent work on acoustic Type I or Type II Weyl points \cite{xiao2015,li2018,ge2018,xie2019}, we design a chiral phononic crystal following the stacking-up approach. Each layer of the structure can be viewed as a 2D Lieb lattice that carries a Dirac point, which is intersecting with a flat band. By introducing the appropriate interlayer coupling, we can construct the triply degenerate spin-$1$ Weyl point in the 3D reciprocal space. 

Figure \ref{fig:FiG2}(a) shows the unit cell of the structure with a lattice constant $a = 20$ mm in $xy$ plane. The unit cell consists of a scattering pillar and a perforated plate base with slanted air tubes. Figure \ref{fig:FiG2}(b) is a top view of the upper pillar, which contains a cylinder {(radius = $\sqrt{\frac{b^2+t^2}{2}}$)} in the center and four wings. The hollow channels (filled by air) between the neighboring solid pillars form the in-plane acoustic wave guide in analogue to the 2D Lieb phononic lattice. The wings have width $L = 10$ mm and thickness $t = 2$ mm. The distance between the centerlines of the two parallel wings is $b=5$ mm. {The colored letters (A, B and C) give us a rough mapping relations between the in-plane wave guides and the tight-binding model shown in {Fig.\ref{fig:FiG1}(a)}}. Figure \ref{fig:FiG2}(c) shows the perforated plate base. The four purple regions represent the slanted air tubes that bring the chiral interlayer couplings. These holes are generated by sweeping a surface [shown in the Fig. \ref{fig:FiG2}(c) inset] in a spiral manner for $90$ degrees. The details of the intralayer and interlayer waveguides are demonstrated in Appendix A. Both the upper pillar and the base plate are of height {$h_p = h_b = 5$} mm, such that the unit cell is of height $H=10$ mm in total in the \textit{z} direction. 

The distribution of the spin-$1$ Weyl points in the first Brillouin zone is shown in Fig. \ref{fig:FiG2}(d). As we can see, there exists a spin-1 Weyl point with topological charge $+2$ at the M point on the $k_z=0$ plane and a spin-$1$ Weyl point carrying topological charge $-2$ located at the $R$ point on the $k_z= \pm \pi/H$ planes. The topological charges of the Weyl points are numerically calculated by the Wilson Loop method \cite{yu2011} (See Appendix B for details). Considering the Weyl points at the $M$ ($R$) points are shared by $4$ ($8$) neighboring Brillouin zones, there are total one pair of spin-$1$ Weyl points with opposite charges ($ \pm 2$) existing in the first Brillouin zone. 

To confirm the existence of the spin-$1$ Weyl points, {we conduct numerical simulations of the acoustic wave dispersions  using the commercial finite element analysis (FEA) software COMSOL.} We show the computational results of the frequency band structures of the unit cell on 2D reduced reciprocal plane for fixed $k_z=0$, $k_z=\pi/2H$, and $k_z=\pi/H$ in Fig. \ref{fig:FiG2}(e). {For comparison, we also include the results based on the long-range tight-binding model (dashed curves), which are in excellent agreement with the FEA results. See Appendix C for details, including the improvement of the tight-binding model's accuracy by implementing long-range model over the short-range one.}

In Fig. \ref{fig:FiG2}(e), we observe that the first three bands of the bulk dispersion diagram are degenerated at the $M$ and $R$ points. While $k_z$ is different from $0$ and $\pm \pi$, the degeneracy at the Weyl points is broken by the synthetic gauge flux introduced by the chiral interlayer couplings. Such trend can be observed through the unit cell band structure along the $MR$ line at the boundary of the first Brillouin zone [see Fig. \ref{fig:FiG2}(f)]. When we fix $k_z=\pi/2H$, the degeneracy is lifted, and two band gaps merge between the first three bands. By evaluating the rotational symmetry of the eigenmodes at the high symmetric points in the reduced 2D Brillouin zone \cite{fang2012}, we know that the two band gaps are both of nonzero Chern number ($-1$), which indicates that they are topologically nontrivial. Based on the bulk-edge correspondence of topology, we expect to see topologically protected localized modes at the boundary of the system.

\begin{figure}[t]
\includegraphics[width=3.5in]{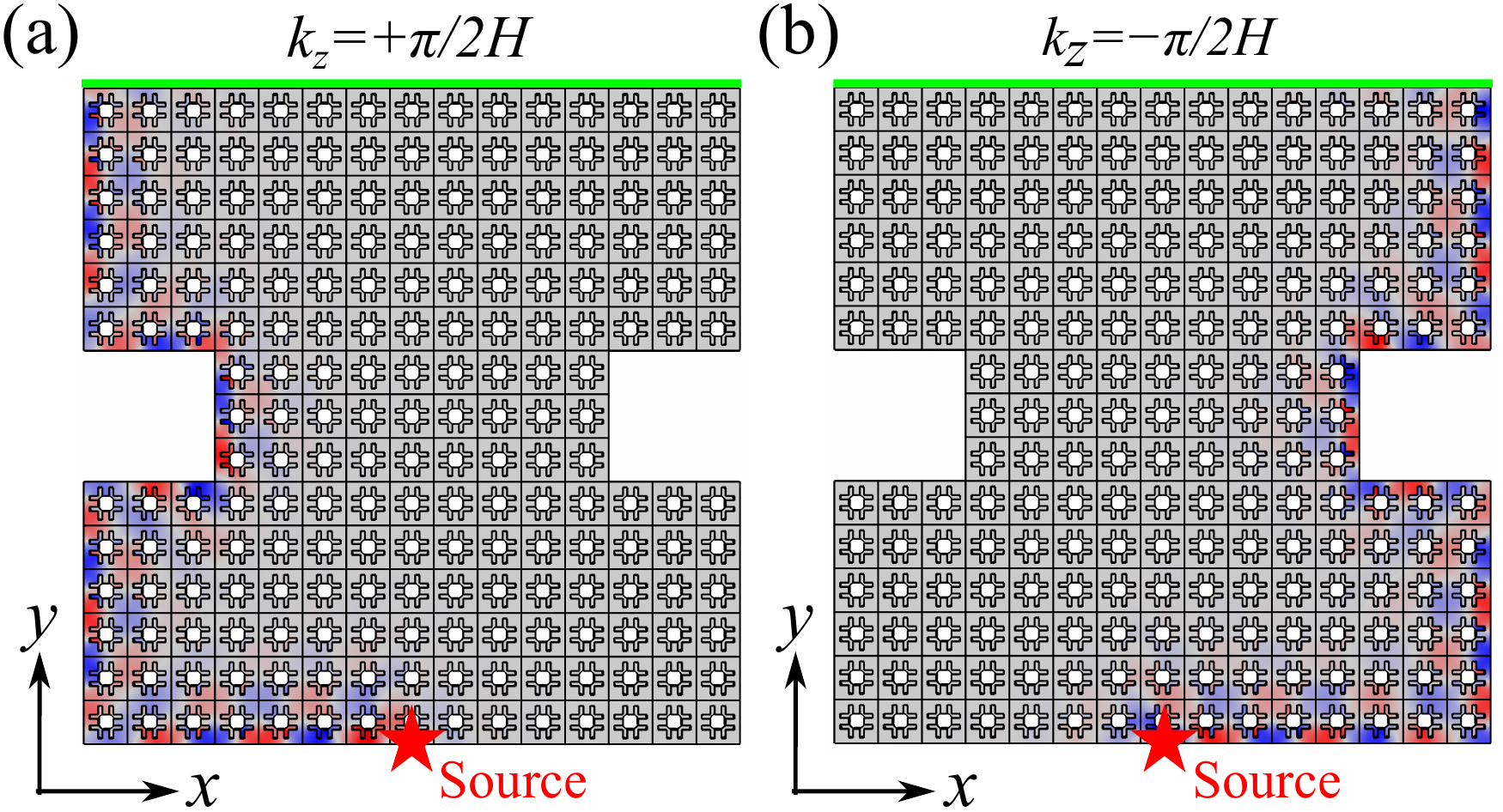}
\caption{One way propagation of the topologically protected surface arc states at $f = 7.8$ kHz for (a) $k_z = + \pi/2H$ and (b) $k_z = - \pi/2H$. Color intensity represents the magnitude of acoustic pressure field.}
\label{fig:FiG4}
\end{figure}

To demonstrate the topologically protected, directional surface arc states in the system, we construct a supercell consisting of $20$ unit cells. The strip is finite in the $y$ direction and has hard boundary conditions on the positive and negative $y$ ends. We apply period boundary conditions in both $x$ an $z$ directions. By fixing $k_z=\pi/2H$ and varying $k_x$ from $-\pi/a$ to $\pi/a$, we obtain the projected band structures in the $x$ direction, as shown in Fig. \ref{fig:FiG3}(a). We can observe that surface arc states emerge in both of the nontrivial band gaps. The red curves represent the modes that are localized at the bottom ends, while the blue curves stand for the modes localized at the top ends. Four localized eigenmodes corresponding to the stars in Fig. \ref{fig:FiG3}(a) are plotted in Fig. \ref{fig:FiG3}(b). By looking at the slope of the bands that represent the surface arc states, we can determine the sign of their group velocities. We know that the top-end (bottom-end) modes will propagate in the positive (negative) $x$ direction. Then the Fermi arc can be obtained by looking at the equifrequency contour of the band structure of the supercell in the 2D Brillouin zone spanned by $k_x$ and $k_z$. Here, the fixed frequency works as an equivalent \textit{Fermi energy} in the acoustic systems. 

The equifrequency contour at $f = 7.8$ kHz (lying in the first band gap) and $f = 10$ kHz (lying in the second band gap) are plotted in  Figs. \ref{fig:FiG3}(c) and (d) respectively. The colored lines represent the acoustic Fermi arcs that connect two Weyl points with opposite charges. Specifically, the red (blue) lines stand for the surface Fermi arcs on the negative (positive) $XZ$ plane. The grey regions represent the projected bulk bands. As we can see, Fermi arcs exist in both band gap regions implying that our structure can support unidirectional surface states in multiple frequency bands. Also the nearly straight Fermi arcs suggest that the wave packages have group velocity parallel to the $x$ directions, which forms the collimation effect of the surface waves.

We then construct an infinite system to verify the robustness of the surface arc states against the defect and sharp bend (Fig. \ref{fig:FiG4}). Such an infinite structure is periodic in the $z$ direction and has finite boundaries in both $x$ and $y$ directions. We introduce a $3 \times 3$ defect on both right and left boundaries. The top boundary is set to be radiative so that the sound waves can leak out to the outside environment, as marked by the green edge in Fig. \ref{fig:FiG4}.  A point source is located at the center of the bottom boundary denoted by the red stars. For fixed $k_z=\pi/2H$ and $f = 7.8$ kHz, we can see that the surface waves only travel in the clockwise direction and can pass around the defect and right angle corner without back-scattering [See Fig. \ref{fig:FiG4}(a)]. If we set $k_z=-\pi/2H$, the surface waves will travel to the opposite direction [see Fig. \ref{fig:FiG4}(b)]. Similarly, the surface arc states at $f = 10$ kHz are demonstrated and confirmed in the Appendix D.

In conclusion, a three-band tight-binding model of a 3D Lieb lattice is introduced to predict the existence of the the spin-$1$ Weyl points. Guided by the tight-binding model, we designed a 3D chiral phononic crystal that carries spin-$1$ Weyl points with topological charge $\pm 2$ in the first Brillouin zone. We observed a special straight type acoustic Fermi arcs and the collimated robust propagation of topological surface arc states in the system. {The key points of this work can be summarized as:

1. We propose the first tight-binding model of 3D Lieb lattice with chiral interlayer hopping. In addition, we include both the short- and long-range hopping terms, which provides deeper understanding of the spin-1 Weyl points in the physics aspect.

2. While most of the existing structures supporting Weyl points and surface arc states are based on the woodpile- or graphene-based design, our design explores a new platform consisting of square shape unit cells to study Weyl physics in the acoustic systems. Moreover, the unit cell is of simple geometry and is designed with ease of assembly . 

3. This study reports the first dual-band topologically protected and collimated surface waves in the spin-1 Weyl structure, which has not been studied before.}

The present results paved new way for manipulating acoustic waves in the 3D structure, which can be potentially extended to other artificial systems of photonic lattices \cite{mei2012} and mechanical lattices \cite{zhu2017a}. 


\begin{acknowledgments}
We thank Dr. Jiun-Haw Chu at the University of Washington for fruitful discussions. X. S., and J. Y. are grateful for the financial support from the U.S. National Science Foundation (CAREER1553202 and EFRI-1741685). 
\end{acknowledgments}

\section*{APPENDIX A: Views of the filling air in the unit cell}
\begin{figure}[h]
	\includegraphics[width=3.4in]{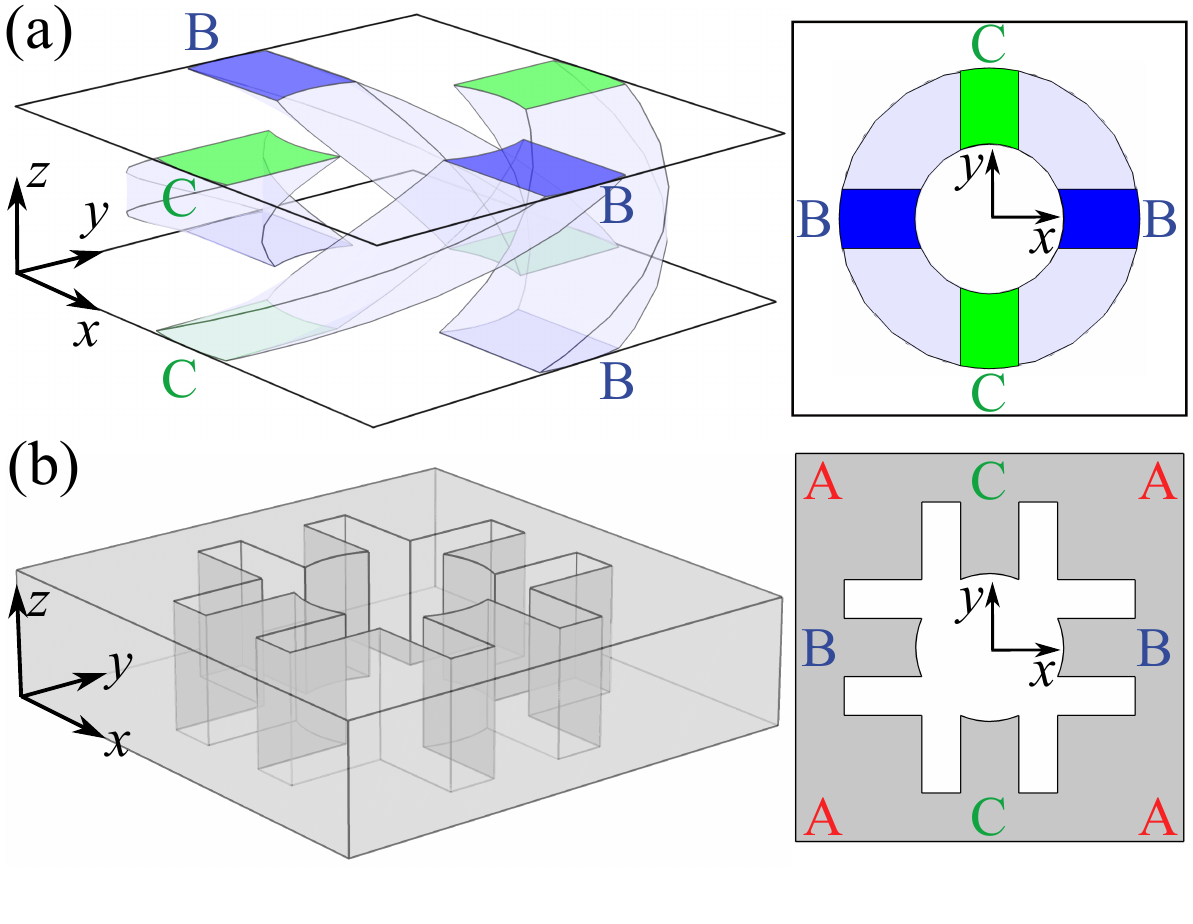}
	\caption{{Oblique and top views of (a) Chiral interlayer air channels and (b) Intralayer acoustic wave guide. The colored letters and surfaces demonstrate the mapping relations between the real phononic crystal unit cell and the effective tight-binding model. In panel (b), the grey area represents the in-plane acoustic wave guide, while the white area indicates the solid pillar.}}
\label{fig:FIGS1}
\end{figure}

In this section, we provide more details of the filling air in the unit cell. Figure. \ref{fig:FIGS1}(a) contains the oblique and top views of the chiral interlayer air channels, which introduces the synthetic gauge flux in the system. As we can see, all the top holes rotate $90$ degrees in a spiral manner with respect to the bottom holes. Figure. \ref{fig:FIGS1}(b) shows the in-plane 2D acoustic waveguide which is formed by the air between two neighboring perforated plates. The scattering pillars help to form an effective 2D Lieb lattice. Then, these neighbouring 2D waveguides are coupled by the interlayer air channels, thereby forming a 3D chiral Lieb lattice. 

\section*{APPENDIX B: Calculation of the Chern numbers with Wilson loop method}

\begin{figure}[h]
	\includegraphics[width=3.4in]{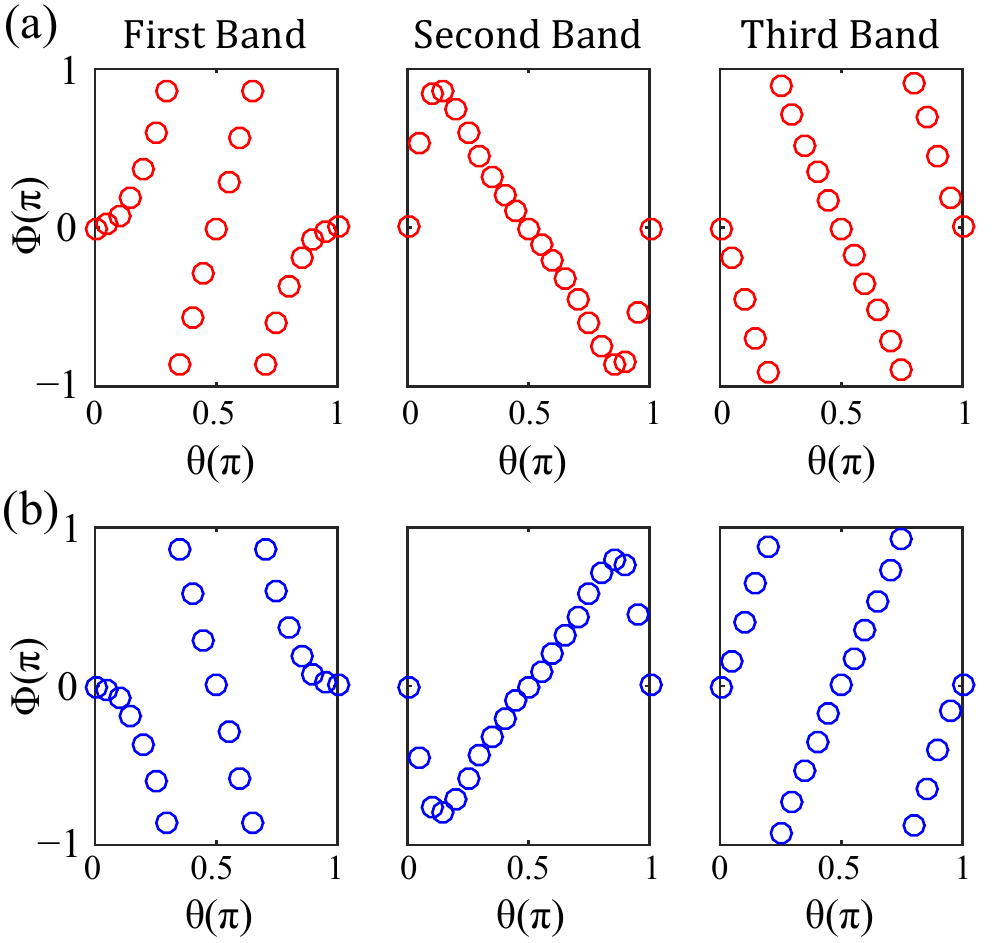}
	\caption{Evolution of the Wannier centers on the spheres enclosing the spin$-1$ point. (a) Wannier centers for the first three bands near the $M$ point. (b) The same for the first three bands near the $R$ point.}
\label{fig:FIGS2}
\end{figure}

The Chern number (or topological charge) can be calculated by integrating Berry curvature on a closed surface enclosing a band degenerate point. Following the method in Refs. \cite{soluyanov2015a,wang2017,yang2019}, we numerically determine the Chern number by tracking the evolution of the Wannier centers on a sphere surrounding the spin$-1$ Weyl point using the Wilson loop method \cite{yu2011}. The calculated Wannier centers on the horizontal loops varing from the north pole to the south pole of the enclosing sphere are presented in Fig. \ref{fig:FIGS2}. As we can see in Fig. \ref{fig:FIGS2}(a), the Wannier centers of the spin$-1$ Weyl point at the $M$ point for the first, second and third bands shift by $+4\pi$, $0$, and $-4\pi$, respectively. This implies that such spin$-1$ Weyl point has a positive charge of $+2$. Similarly, by looking at Fig. \ref{fig:FIGS2}(b), we can conclude that there exist a spin$-1$ Weyl point of charge $-2$ located at $R$ point in the first Brillouin zone.

\section*{APPENDIX C: Unit cell band structure calculated by the tight-binding model and full wave simulations}

\begin{figure}[h]
	\includegraphics[width=3.4in]{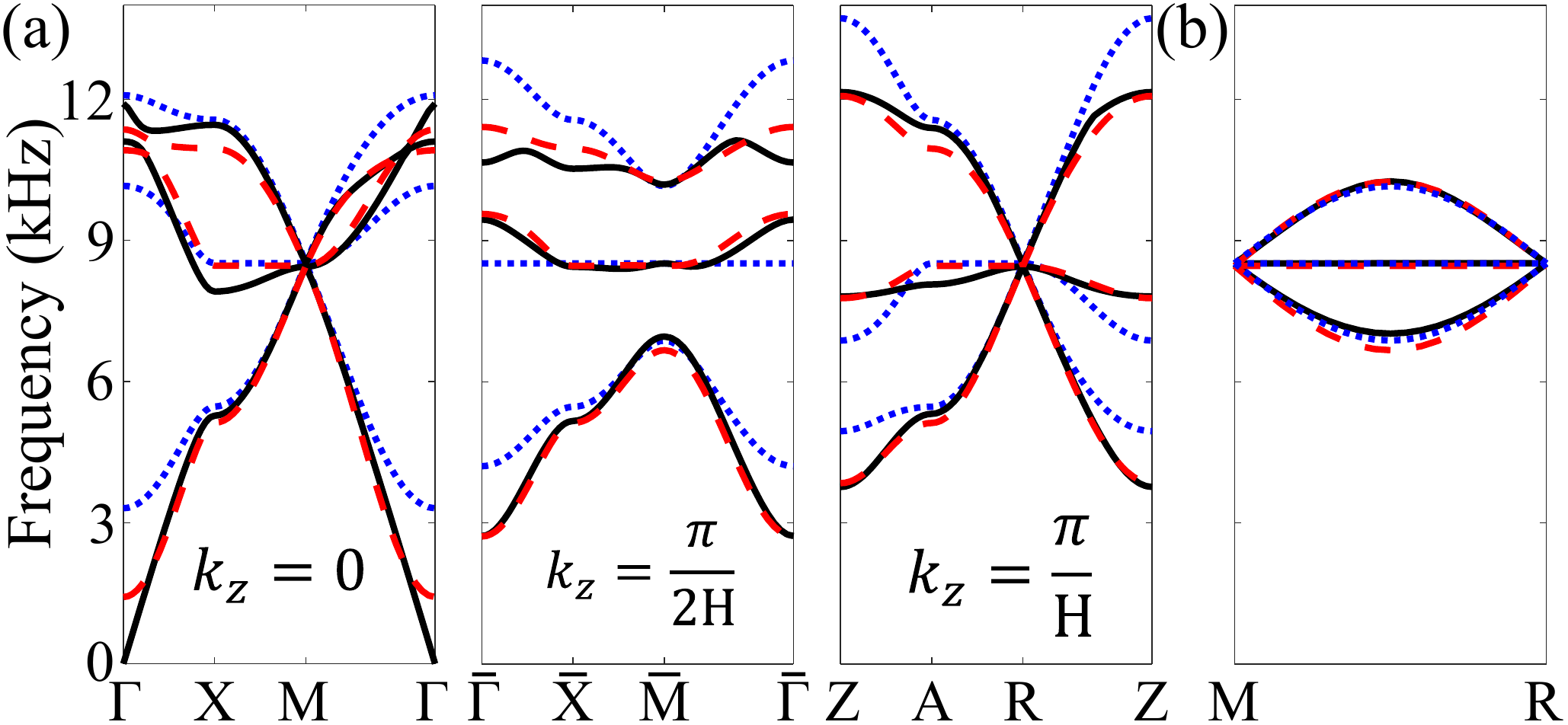}
	\caption{Bulk band structure of the unit cell obtained by tight-binding models and full wave simulations in the reduced 2D reciprocal $k_x-k_y$ planes and $K-H$ lines in the first Brillouin zone.}
\label{fig:FIGS3}
\end{figure}

{In Fig. {\ref{fig:FIGS3}}, we compare the frequency band structures of the unit cell obtained by tight-binding models and the full wave simulations using COMSOL. The solid black lines are the results of COMSOL simulations, which is the same as those ploted in Fig. {\ref{fig:FiG2}} (e) and (f). The results of simplified (i.e., short-range model with $t_{n2}=t_{n3}=t_{n4}=0$) and full three-band (i.e., long-range) tight-binding model are shown as the blue dot lines and red dashed lines respectively. The hopping parameters used in a tight-binding model of both cases are determined by fitting the results with the results of COMSOL full wave simulations. As we can see in Fig. {\ref{fig:FIGS3}}, the simplified Hamiltonian (blue dot line) only works within a small range near the spin-1 Weyl point with the fitting parameters given as $\varepsilon_1=\varepsilon_2=8.516$, $t_{n1}=-1.524$ and $t_{c}=-0.410$. However, by taking into account more hopping terms, the full tight-binding model (red dashed line) can capture the band structures of real phononic crystal very well in the whole Brillouin zone. Here, we set $\varepsilon_1=8.5448$, $\varepsilon_2=7.9962$, $t_{n1}=-1.436$, $t_{n2}=-0.509$, $t_{n3}=-0.232$, $t_{n4}=0.021$, $t_{c}=-0.448$ in the full tight-binding model calculation.}

\section*{APPENDIX D: Surface states at $f=10$ kHz}

\begin{figure}[h]
	\includegraphics[width=3.4in]{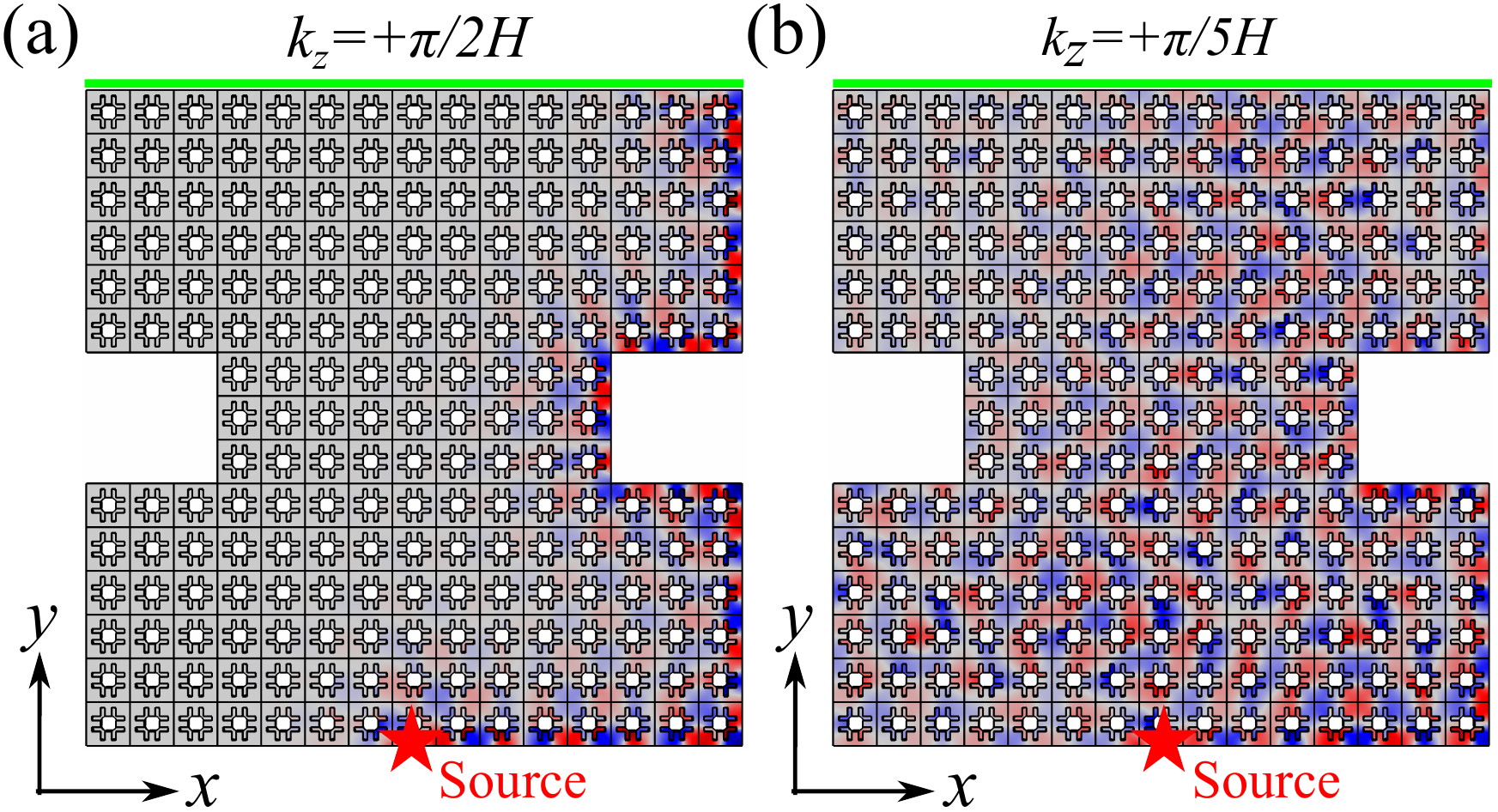}
	\caption{Surface arc states at $f = 10$ kHz which lies in the second topological nontrivial band gap for (a) $k_z = + \pi/2H$ and (b) $k_z = + \pi/5H$. Color intensity represents the magnitude of acoustic pressure field.}
\label{fig:FIGS4}
\end{figure}

In the main manuscript, we show the equifrequency contours of supercell at $f=10$ kHz in Fig. 3(d). It is clear that, for several $k_z$ ranges, the structure support bulk modes and surface modes simultaneously. In this case, we can hardly excite the clear surface modes since they are coupled with the bulk modes and will easily leak to the bulk. To demonstrate such effect, we conduct numerical simulations under surface excitation at $f=10$ kHz for $k_z = + \pi/2H$ or $k_z = + \pi/5H$ [see Fig. \ref{fig:FIGS4}]. Similar to the setup in the main context, the structure is infinite in the $z$ direction with a radiative boundary placed on the top edge, denoted by the green line. For $k_z = + \pi/2H$, the surface states can be clearly observed as all the energy are well confined to the boundary of the structure. However, when we set $k_z = + \pi/5H$, we can barely see the surface waves near the excitation point, and the energy quickly leaks into the bulk of the system.

\bibliography{PRB} 
\bibliographystyle{apsrev}

\end{document}